\documentclass[twocolumn,floatfix,showpacs,preprintnumbers,amsmath,amssymb,superscriptaddress,prb]{revtex4}

\usepackage{graphicx}
\usepackage{dcolumn}
\usepackage{bm}

\begin{document}


\title{Evolution of magnetism in Yb(Rh$_{1-x}$Co$_x$)$_2$Si$_2$}

\author{C. Klingner}
\altaffiliation[Present address: ]{Max-Planck Institute of Biochemistry, D-82152 Martinsried, Germany}
\affiliation{Max-Planck-Institute for Chemical Physics of Solids, D-01187 Dresden, Germany}
\author{C. Krellner}
\email{krellner@cpfs.mpg.de}
\affiliation{Max-Planck-Institute for Chemical Physics of Solids, D-01187 Dresden, Germany}
\author{M. Brando}
\affiliation{Max-Planck-Institute for Chemical Physics of Solids, D-01187 Dresden, Germany}
\author{D.V. Vyalikh}
\author{K. Kummer}
\author{S. Danzenb\"acher}
\author{S.L. Molodtsov}
\author{C. Laubschat}
\affiliation{Institut f\"ur Festk\"orperphysik, Technische Universit\"at Dresden, D-01062 Dresden, Germany}
\author{T. Kinoshita}
\author{Y. Kato}
\author{T. Muro}
\affiliation{Japan Synchrotron Radiation Research Institute, Hyogo 679-5198, Japan}
\author{C. Geibel}
\author{F. Steglich}
\affiliation{Max-Planck-Institute for Chemical Physics of Solids, D-01187 Dresden, Germany}
\date{\today}

\begin{abstract}
We present a study of the evolution of magnetism from the quantum critical system YbRh$_2$Si$_2$ to the stable trivalent Yb system YbCo$_2$Si$_2$. Single crystals of Yb(Rh$_{1-x}$Co$_x$)$_2$Si$_2$ were grown for $0 \leq x \leq 1$ and studied by means of magnetic susceptibility, electrical resistivity, and specific heat measurements, as well as photoemission spectroscopy. The results evidence a complex magnetic phase diagram, with a non-monotonic evolution of $T_N$ and two successive transitions for some compositions resulting in two tricritical points. The strong similarity with the phase diagram of YbRh$_2$Si$_2$ under pressure indicates that Co substitution basically corresponds to the application of positive chemical pressure. Analysis of the data proves a strong reduction of the Kondo temperature $T_K$ with increasing Co content, $T_K$ becoming smaller than $T_N$ for $x\approx 0.5$, implying a strong localization of the $4f$ electrons. Furthermore, low-temperature susceptibility data confirm a competition between ferromagnetic and antiferromagnetic exchange. The series Yb(Rh$_{1-x}$Co$_x$)$_2$Si$_2$ provides an excellent experimental opportunity to gain a deeper understanding of the magnetism at the quantum critical point in the vicinity of YbRh$_2$Si$_2$ where the antiferromagnetic phase disappears ($T_N\rightarrow 0$). 
\end{abstract}

\maketitle

\section{\label{sec:Introduction} Introduction}
In recent years the heavy fermion (HF) metal YbRh$_2$Si$_2$ has been intensively investigated due to its proximity to an antiferromagnetic (AFM) field-induced quantum critical point (QCP)\cite{custers_break-up_2003}. This proximity leads to extraordinary non-Fermi-liquid (NFL) behavior, such as the divergence of the electronic Sommerfeld coefficient, \cite{custers_break-up_2003, gegenwart_magnetic-field_2002, oeschler_low-temperature_2008} the linear-in-$T$ resistivity, \cite{trovarelli_ybrh2si2_2000} as well as diverging Gr\"uneisen ratios. \cite{kuechler_divergence_2003,tokiwa_divergence_2009} The reason of the NFL behavior is assumed to result from quantum fluctuations with a strong ferromagnetic (FM) component\cite{ishida_ybrh2si2:_2002, gegenwart_ferromagnetic_2005}. Moreover, a jump of the Fermi volume has been observed in Hall-effect\cite{paschen_hall-effect_2004, FriedemannPNAS} and thermopower\cite{Hartmann:2010} measurements, which could not be explained by the conventional 3D spin-density-wave (SDW) scenario.\cite{gegenwart_quantum_2008}  The recent discoveries of an additional energy scale vanishing at the QCP, which does neither correspond to the N\'eel temperature nor to the upper boundary of the Fermi-liquid region  \cite{Gegenwart:2007}, and a large critical exponent $\alpha=0.38$ at the AFM phase transition observed in low-temperature specific-heat measurements on a single crystal of superior quality \cite{Krellner:2009} have once again boosted the interest in YbRh$_2$Si$_2$. Theoretically, the unconventional quantum criticality in this material may be ascribed to a breakdown of the Kondo screening at the field induced AFM QCP. \cite{si_locally_2001,coleman_what_2002, Senthil:2004}

Besides strong experimental efforts, the determination of the magnetic structure using neutron scattering experiments was not yet successful due to the small size of the ordered moment ($\approx 10^{-3}\,\mu_B$)\cite{Ishida:2003} and the very low N\'eel temperature $T_N=72$\,mK. One way to overcome these difficulties would be the application of hydrostatic pressure, which results in a stabilization of the AFM ordering and, therefore, in an increase of the ordered moment and $T_{\rm{N}}$. \cite{Mederle:2002} In addition, the appearance of a second phase transition, labeled $T_{\rm{L}}$, has been reported at $p\geq 1$\,GPa.\cite{Mederle:2002} An alternative to hydrostatic pressure is isoelectronic doping with cobalt which leads to chemical pressure because of a smaller unit-cell volume, $V$. Hodges already showed earlier that YbCo$_2$Si$_2$ orders antiferromagnetically at $T_N=1.7$\,K with a sizable ordered Yb moment of $1.4\,\mu_B$.\cite{hodges_magnetic_1987} Later on, susceptibility and inelastic neutron scattering experiments confirmed a stable trivalent Yb state with well defined crystal electric field levels.\cite{kolenda_valence_1989} Several results on RCo$_2$Si$_2$ compounds proved Co to be non-magnetic, because of a strong Co-Si hybridization.\cite{Szytula:1989} First experiments with small amounts of Co doping led already to very promising results,\cite{Westerkamp:2008,Friedemann:2009} when compared to the equivalent hydrostatic pressure experiments on YbRh$_2$Si$_2$.\cite{Mederle:2002,Plessel:2003,Knebel:2006} We therefore studied the complete doping series Yb(Rh$_{1-x}$Co$_x$)$_2$Si$_2$ in order to get a reliable picture of the evolution and the characteristics of the magnetic order. 

The paper is organized in the following way. In the first section the crystallographic analysis of all doped samples are given, followed by the presentation of the magnetic susceptibility results. The subsequent sections are dedicated to the electrical-resistivity data followed by the presentation of the specific-heat results. Finally, the experimental results of the photoemission spectroscopy are shown. From these data we derive the magnetic phase diagram of the complete series Yb(Rh$_{1-x}$Co$_x$)$_2$Si$_2$. Its close resemblance to the magnetic phase diagram of YbRh$_2$Si$_2$ under pressure confirms that the present alloy system provides an excellent access to a deeper understanding of the unusual properties of YbRh$_2$Si$_2$.
 
\section{\label{sec:Experimental} Experimental results}

\subsection{\label{sec:Crystallographics}Crystallographic parameters}

\begin{figure}
\includegraphics[width=\columnwidth]{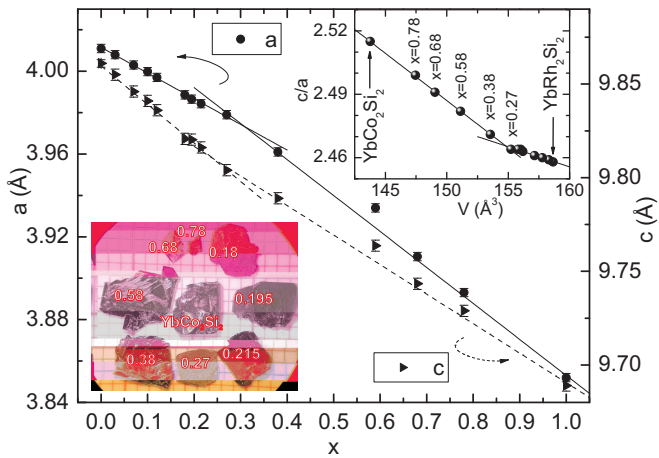}%
\caption{\label{FigXRD} (Color online) Lattice parameters $a$ and $c$ of the Yb(Rh$_{1-x}$Co$_x$)$_2$Si$_2$ series. The slope of the data indicates positive chemical pressure with increasing $x$. Left inset: Picture of typical single crystals for all Co concentrations. Right inset: $c/a$ ratio versus unit cell volume. The lines represents linear fits to the experimental data. }%
\end{figure}

Single crystals of Yb(Rh$_{1-x}$Co$_x$)$_2$Si$_2$ were grown using the indium flux-growth technique developed for YbRh$_2$Si$_2$\cite{krellner_yb_2009}. Surprisingly, the growth parameters optimized for the pure Rh-based compound could be retained for the whole alloy series, enabling us to grow large single crystals up to pure YbCo$_2$Si$_2$ (see left inset, Fig.~\ref{FigXRD}). All samples were prepared in Al$_2$O$_3$-crucibles, which were placed in a second crucible of tantalum closed under argon atmosphere using arc melting.  The elements were heated up to $T=1500^{\circ}$\,C, followed by a slow cooling to $T=1000^{\circ}$\,C. After the growth, the single crystals were retrieved from the In-flux using hydrochloric acid for chemical etching. None of the crystals showed any indication of surface destruction due to HCl attack. Only for the Co concentration $x=0.27$ the surface  of the crystals appeared to be a little different, probably due to residue elements after the etching process (see left inset, Fig.~\ref{FigXRD}). However, after polishing the surface, different physical measurements led to the conclusion that the bulk properties were not affected.

To find out the crystal orientation, the single crystals were analyzed using Laue backscattering. The Laue-reflection pictures confirmed that the surface of the crystal is always perpendicular to the crystallographic $c$-axis. Therefore, the crystal growth in the $ab$-direction is favored with respect to the $c$-direction. Subsequently, X-ray powder-diffraction patterns on crushed single crystals showed that the complete doping series is single phase with the expected ThCr$_2$Si$_2$-structure (space group I4/mmm), in agreement with the crystal structure of YbRh$_2$Si$_2$ and YbCo$_2$Si$_2$.\cite{Rossi:1979} Since one aim of this work is to compare the effects of hydrostatic and chemical pressure, the cell parameters $a$ and $c$ should both decrease with increasing doping. Otherwise, the comparison would be more difficult due to a non-isotropic change of the unit cell. Fortunately, for the complete series Yb(Rh$_{1-x}$Co$_x$)$_2$Si$_2$ both lattice parameters $a$ and $c$ continuously decrease (Fig.~\ref{FigXRD} and Tab.~\ref{TabXRD}). The ratio $c/a$ shows only a weak linear increase with $x$ with different slopes below and above $x\approx 0.38$, the former being very small. Since hydrostatic pressure experiments evidence a constant $c/a$ up to highest investigated pressure of $p=21$\,GPa,\cite{Plessel:2003} this slight change of $c/a$ has to be taken into account in the comparison of hydrostatic and chemical pressure results.

The Co content was determined using energy dispersive X-ray spectra performed on a scanning electron microscope (Philips XL30) with a Si(Li)-X-ray detector. Three crystals of each batch with a polished surface were analyzed at several positions. The respective mean values result in the given percentages with a relative error of less than $1\,\rm{at.}\%$. These real Co concentrations, which are used for $x$ throughout this paper, are somewhat larger than the nominal concentrations due to different solubilities of the elements in the In-flux.
 
\begin{table}
\caption{\label{TabXRD}Lattice parameters for the complete series obtained from X-ray analysis and least square root fitting procedures.}
\begin{ruledtabular}
\begin{tabular}{dcccc}
\multicolumn{1}{c}{x} 		& $a$ (\AA) 		& $c$ (\AA) & $c/a$	& $V$ (\AA$^3$)\\
\pm 0.01 & 	$\pm0.001$		& $\pm0.002$	& 	$\pm0.001$		& $\pm0.1$			\\\hline  
0.0 	& $4.011$		& $9.861$ & $2.458$ & $158.6$\\
0.03 	& $4.008$		& $9.855$ & $2.459$	& $158.3$\\
0.07 	& $4.003$		& $9.846$ &	$2.460$	& $157.8$\\
0.12	& $3.997$		& $9.836$ &	$2.461$	& $157.1$\\
0.18  	& $3.988$		& $9.821$ &	$2.463$	& $156.2$\\
0.195	& $3.986$		& $9.820$ &	$2.464$	& $156.0$\\
0.215	& $3.984$		& $9.816$ &	$2.464$	& $155.8$\\
0.27	& $3.979$		& $9.804$ &	$2.464$	& $155.2$\\
0.38	& $3.961$		& $9.787$ &	$2.471$	& $153.6$\\
0.58	& $3.934$ 		& $9.764$ &	$2.482$	& $151.1$\\
0.68	& $3.911$		& $9.743$ &	$2.491$	& $149.0$\\
0.78	& $3.893$		& $9.729$ &	$2.499$	& $147.5$\\
1.0		& $3.852$		& $9.689$ &	$2.515$	& $143.8$\\
\end{tabular}
\end{ruledtabular}
\end{table}

\subsection{\label{Susceptibility} Magnetic susceptibility}

\begin{figure}
\includegraphics[width=\columnwidth]{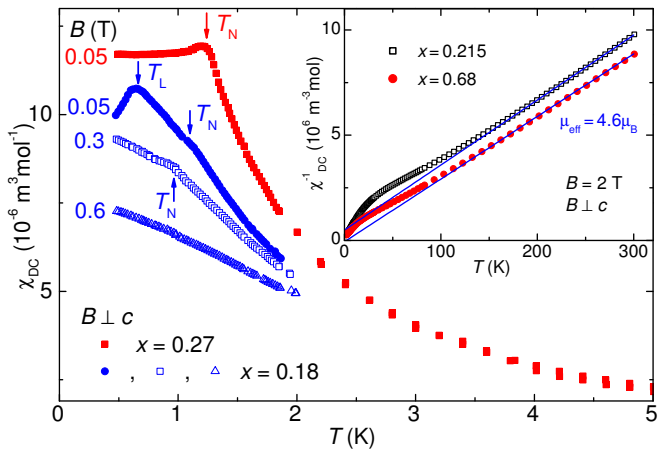}%
\caption{\label{FigChi}(Color online) Temperature dependence of the magnetic susceptibility for $x=0.18$ and $0.27$. The AFM transition temperatures $T_{\rm{L}}$ and $T_{\rm{N}}$ are marked by arrows. The inverse susceptibility is exemplarily shown in the inset for $x=0.215$ and $0.68$ presenting paramagnetic Curie-Weiss behavior above 150\,K  with $\mu_{\rm{eff}}=4.6\mu_{\rm{B}}$. }%
\end{figure}

In this section, we discuss the magnetic behavior of the Yb(Rh$_{1-x}$Co$_x$)$_2$Si$_2$ single crystals. All measurements were carried out using a Magnetic Property Measurement System (MPMS) of Quantum Design. We also present results of the dc-susceptibility down to $T=0.5$\,K, which where realized with a $^3$He-option for the MPMS designed by the iQuantum Corporation. 

Dc-susceptibility ($\chi$) measurements on YbRh$_2$Si$_2$ revealed Curie-Weiss-law behavior between $200\leq T\leq300$\,K with a large magnetic anisotropy.\cite{trovarelli_low-temperature_2000} At high temperatures, the effective moments along $a$ and $c$, $\mu_{\rm eff}^{a}$ and $\mu_{\rm eff}^{c}$, equal the value of $4.54\,\mu_B$ corresponding to the $J=7/2$ multiplet of the free Yb$^{3+}$ ion. Because the strong easy-plane anisotropy is retained throughout the whole alloy, we focus here on the in-plane susceptibility results. In the temperature range $200\leq T\leq 300$\,K, $\chi(T)$ shows Curie-Weiss (CW)-like behavior with an effective moment $\mu_{\rm{eff}}^{ab}=4.6\pm0.1\mu_{\rm{B}}$ for all $x$ (see Tab.~\ref{TabCW} and inset of Fig.~\ref{FigChi}), as expected for a trivalent Yb state, and thus supporting the absence of a Co moment. The difference of the Weiss temperatures, $\Theta_{\rm{W}}^{\rm{ab};200-300\rm{K}}-\Theta_{\rm{W}}^{\rm{c};200-300\rm{K}}$ (see Tab.~\ref{TabCW}), changes only slightly throughout the complete series, indicating a rather smooth variation of the leading crystal electric field (CEF)-parameter $B^0_2$ (Stevens coefficient) when going from YbRh$_2$Si$_2$ to YbCo$_2$Si$_2$. \cite{Boutron:1973, Avila:2004} This is supported by our analysis of the CEF-scheme for $x=1$, \cite{Klingner:2010a} resulting in a $\Gamma_7$ ground state similar to YbRh$_2$Si$_2$. \cite{Vyalikh:2010} 

The first low-temperature ac-susceptibility measurements for $x\leq 0.12$ were performed and reported by Westerkamp et al.\cite{Westerkamp:2008} They already confirm the stabilization of the AFM ordering with increasing $x$. In addition to the increasing $T_{\rm{N}}$,  a second magnetic feature appears in the magnetically ordered region for $x\geq0.07$. Various measurements on YbRh$_2$Si$_2$ under hydrostatic pressure showed the appearance of a second anomaly, $T_{\rm{L}}$, at $p\geq 1$\,GPa.\cite{Mederle:2002,Knebel:2006} Focusing now on the next higher doping $x=0.18$, the two mentioned anomalies are clearly visible in Fig.~\ref{FigChi} for a magnetic field $B=50$\,mT, applied perpendicular to the hard magnetic $c$-axis. At $T_{\rm{N}}$, a small change of slope in the temperature dependence marks the onset of AFM order; whereas, below $T_{\rm{L}}$, the typical decrease of $\chi(T)$ is observed. The fact that the ordering is of AFM type can be deduced from the temperature and field dependence of the magnetic susceptibility, as well as from the shifting of $T_{\rm{N}}$ and $T_{\rm{L}}$ towards lower temperatures with increasing external magnetic fields (cf. Fig.~\ref{FigChi}). For $x=0.18$, an in-plane field of $B=0.3$\,T is sufficient to suppress the phase transition at $T_{\rm{L}}$, while the magnetic ordering has completely disappeared at $B=0.6$\,T. Increasing the doping to $x=0.27$, $T_{\rm{L}}$ is not observable anymore for $T\geq 0.5$\,K, whereas $T_{\rm{N}}$ is shifted towards higher temperatures. The $T$-dependence of $\chi$ is still of AFM type but changes profoundly compared to the one observed for $x=0.18$. Recent resistivity measurements on the $x=0.27$ sample showed that no second phase transition  appears down to $T=20$\,mK.\cite{Lausberg_private}

\begin{figure}%
\includegraphics[width=\columnwidth]{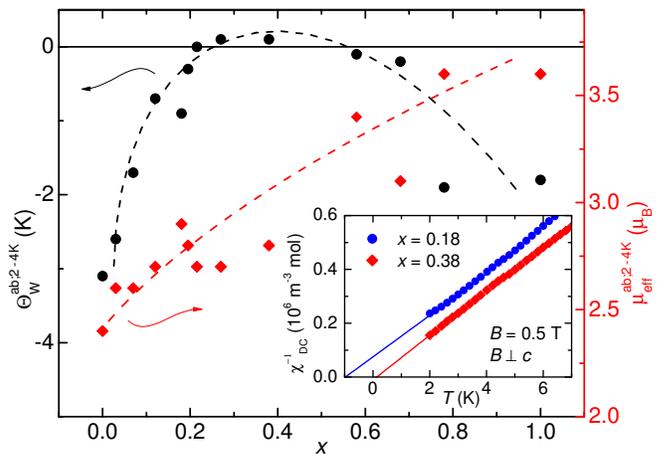}%
\caption{\label{FigCW}(Color online) Low-temperature Weiss temperature, $\Theta_{\rm{W}}^{\rm{ab};2-4\rm{K}}$, for the complete series showing the increasing FM component of the exchange interactions with increasing $x$ (left axis, circles). The effective moments in the same temperature region, $\mu_{\rm{eff}}^{\rm{ab};2-4\rm{K}}$, indicate a decreasing Kondo screening  for higher cobalt contents (right axis, diamonds). Dashed lines are guide to the eyes. The inset shows exemplarily the Curie-Weiss behavior together  with the linear fit for two different concentrations.}%
\end{figure}

\begin{table*}
\caption{\label{TabCW}Effective moment, $\mu_{\rm{eff}}$, Weiss temperature, $\Theta_{\rm{W}}$, from the high-temperature ($200\leq T\leq300$\,K) and the low-temperature ($2\leq T\leq4$\,K) CW fits for both the magnetic easy plane ($ab$) and the hard axis ($c$). The magnetic phase transition temperatures, $T_{\rm{L}}$ and $T_{\rm{N}}$, were obtained from heat-capacity measurements or from the literature; * denote first-order phase transitions.}
\begin{ruledtabular}
\begin{tabular}{dddddddd}
\multicolumn{1}{c}{$x$}	& \multicolumn{1}{c}{$\mu_{\rm{eff}}^{\rm{ab};200-300\rm{K}}(\mu_B)$} & \multicolumn{1}{c}{$\Theta_{\rm{W}}^{\rm{ab};200-300\rm{K}}(\rm K)$} & \multicolumn{1}{c}{$\Theta_{\rm{W}}^{\rm{c};200-300\rm{K}}(\rm K)$} & \multicolumn{1}{c}{$\mu_{\rm{eff}}^{\rm{ab};2-4\rm{K}}(\mu_B)$} & \multicolumn{1}{c}{$\Theta_{\rm{W}}^{\rm{ab};2-4\rm{K}}(\rm K)$}    & \multicolumn{1}{c}{$T_{\rm{L}}(\rm K)$} & \multicolumn{1}{c}{$T_{\rm{N}}(\rm K)$}\\
			&	\pm0.1 & \pm10 & \pm10 & \pm0.2 & \pm0.2 & \pm0.03& \pm0.03\\ \hline
0.0	& 4.5 & -9 & -215 & 2.4 & -3.1 & -	& 0.072\cite{Krellner:2009}\\
0.03 &4.6 &-30 &  - & 2.6 & -2.6 & - & 0.18\cite{Westerkamp:2008}\\
0.07 &4.6 &-43 &  - & 2.6 & -1.7 & 0.06\cite{Westerkamp:2008} & 0.41\cite{Westerkamp:2008}\\
0.12 &4.5 &-31 & -185 & 2.7 & -0.7 & 0.28\cite{Westerkamp:2008}&0.73\cite{Westerkamp:2008}\\
0.18 &4.6 & -23 & - & 2.9 & -0.9 & 0.65 & 1.10\\
0.195&4.7 & -22 & - & 2.8 & -0.3 & 0.74 & 1.11\\
0.215&4.6 & -16 & - & 2.7 & 0.0  & 0.91^* &1.14\\
0.27 &4.6 & -11 & -155 & 2.7 & 0.1  &  -     & 1.30\\
0.38 &4.7 & -10 & - & 2.8 & 0.1  & -      & 1.22\\
0.58 & -    &   -   & - & 3.4 & -0.1 & -      & 0.73\\
0.68 &4.6 & 3   & - & 3.1 & -0.2 & 1.06^* &1.14\\
0.78 & -    &   -   & - & 3.6 & -1.9 & 1.01^* &1.32\\
1.0    &4.7 & -4  & -160& 3.6 &-1.8 & 0.91^*&1.61\\			
\end{tabular}
\end{ruledtabular}
\end{table*}

The important role of the FM fluctuations in YbRh$_2$Si$_2$ has been already mentioned. As pointed out by Ishida \textit{et al.}\cite{ishida_ybrh2si2:_2002} for YbRh$_2$Si$_2$ and by Gegenwart \textit{et al.}\cite{gegenwart_ferromagnetic_2005} for YbRh$_2($Si$_{0.95}$Ge$_{0.05})$, the quantum critical fluctuations have a very strong FM component. Among all other known quantum critical HF compounds it is also this feature which makes the HF system YbRh$_2$Si$_2$ unique. In the following, we discuss the evolution of the FM fluctuations above $T_{\rm{N}}$ upon increasing the chemical pressure. In systems with a stable Yb$^{3+}$ state, because of the small de-Gennes factor, magnetic inter-site exchanges are much weaker ($\lesssim 10$\,K) than crystal field effects ($\lesssim 100$\,K). Then, the Weiss temperature obtained from a fit to the low-temperature susceptibility data reflects the sum of all exchange interactions (while the fit to the high-temperature data reflects the CEF effect) and is a good indicator for the dominant exchange interactions. We therefore fitted the measured inverse susceptibility at $B=0.5$\,T with a CW law between $2\leq T\leq 4$\,K, just above $T_{\rm{N}}$ (see inset of Fig.~\ref{FigCW}) for the complete series. The values of $\Theta_{\rm{W}}^{\rm{ab};2-4\rm{K}}$ and $\mu_{\rm{eff}}^{\rm{ab};2-4\rm{K}}$ deduced from this fit are given in Tab.~\ref{TabCW}.  They are also plotted in Fig.~\ref{FigCW} to show the overall evolution. It is clearly visible that $\Theta_{\rm{W}}^{\rm{ab};2-4\rm{K}}$ (circles) is increasing monotonically with increasing $x$ for $x\leq 0.38$ and even changes sign at $x\approx 0.2$. This leads to the important result that, with increasing $x$, the strength of the FM fluctuations increases and even dominate for $0.27\leq x\leq0.38$ ($\Theta_{\rm{W}}^{\rm{ab};2-4\rm{K}}>0$), although all samples of the series order antiferromagnetically at low temperatures. For the stoichiometric YbCo$_2$Si$_2$ compound the Weiss temperature is again negative, $\Theta_{\rm{W}}^{\rm{ab};2-4\rm{K}}=(-1.8\pm0.2)$\,K.

The second fit parameter of the low-temperature CW fit, $\mu_{\rm{eff}}^{\rm{ab};2-4\rm{K}}$, reflects the evolution of the Kondo screening. In Yb systems, increasing chemical pressure should lead to a decreasing Kondo temperature and, therefore, to a change in the dominating exchange interaction from Kondo to Ruderman-Kittel-Kasuya-Yosida (RKKY). In Fig.~\ref{FigCW}, the diamonds show the values of $\mu_{\rm{eff}}^{\rm{ab};2-4\rm{K}}$ for all $x$. The monotonic increase of $\mu_{\rm{eff}}^{\rm{ab};2-4\rm{K}}$ from 2.4 to 3.6 $\mu_B$ supports a decrease of the Kondo screening with increasing $x$, resulting in a larger effective moment. The final discussion about the interplay between Kondo effect and RKKY exchange interaction will be given in Sec.\,III.B.

\subsection{\label{SecResistivity}Electrical resistivity}
\begin{figure}%
\includegraphics[width=\columnwidth]{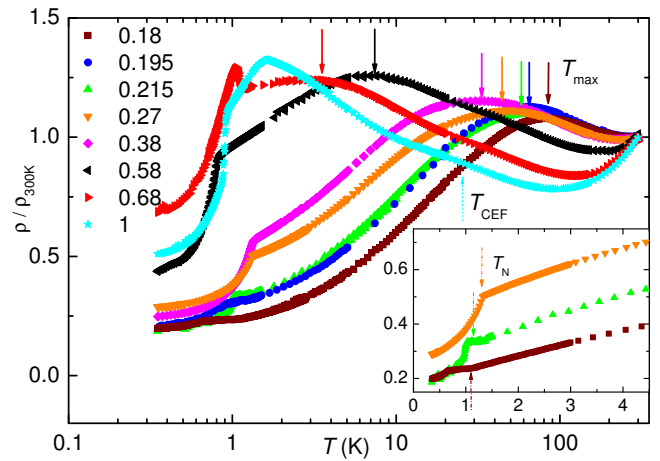}%
\caption{\label{FigRho}(Color online) Temperature dependence of the resistivity normalized to the room-temperature value for $0.18\leq x\leq 1$. The downward arrows mark the onset of coherence effect typical for Kondo lattice systems. With increasing $x$, $T_{\rm{max}}$ shifts towards lower temperatures. The dashed arrow shows the CEF related hump for $x\geq 0.58$. In the inset, the low-temperature behavior of the resistivity for the Co concentrations $0.18\leq x\leq 0.27$ is shown. Dash-dotted arrows indicate $T_{\rm{N}}$}%
\end{figure}

In this section, we present the results of electrical resistivity measurements on single crystals with $0.18\leq x\leq1$. The resistivity $\rho(T)$ was measured between $0.35\leq T\leq300$\,K using the four-point contact method and the ac-transport option of a Physical Property Measurement System (PPMS) from Quantum Design, equipped with a $^3$He-insert. All samples were measured with the current perpendicular to the crystallographic $c$-direction and were normalized to their room-temperature value, $\rho(300\,\rm{K})$, for comparison. In Fig.~\ref{FigRho}, this normalized resistivity  is plotted on a logarithmic $T$-scale. For $0.18\leq x\leq 0.68$, a maximum is clearly visible (arrows in Fig.~\ref{FigRho} are indicating the positions of the maximum, $T_{\rm{max}}$). This results from the interplay between two different energy scales: (1) the CEF splitting and (2) the onset of the coherence effect related to the Kondo energy scale, $T_{\rm{K}}$, as observed in various other Kondo lattices.\cite{gupta_theoretical_1988} Since the susceptibility measurements show that the CEF does not change drastically with increasing chemical pressure, the shift of $T_{\rm{max}}$ with $x$ is unlikely to result from a change of the CEF splitting. Furthermore, in previous pressure experiments on pure YbRh$_2$Si$_2$, $T_{\rm{max}}$ also decreases with increasing pressure. \cite{Plessel:2003} We, thus, conclude that the reduction of $T_{\rm{max}}$ for increasing $x$ results from the decrease of $T_{\rm{K}}$. A weak and broad hump in $\rho(T)$ emerges around 25\,K (dashed arrow in Fig.~\ref{FigRho}) for $x \geq 0.58$, which  can be related to the CEF splitting. This clearly marks the detaching of the two energy scales, $T_{\rm{K}}$ and  CEF effects, in $\rho(T)$. For $x=1$, $T_{\rm{max}}$ is not any more distinguishable because the AFM order sets in before the coherence effect can develop. This leads to the result that at some point in the doping series the energy scales of the two interactions, Kondo and RKKY, must cross each other and the system changes from a more itinerant Kondo lattice to a magnetically localized system. A detailed discussion on this interplay will be given in Sec.\,III.B.

Now, we will concentrate on the low-temperature behavior around the AFM ordering. For all concentrations the onset of the antiferromagnetism at $T_{\rm{N}}$ can be deduced from an anomaly in $\rho(T)$ (see dash-dotted arrows in the inset of Fig.~\ref{FigRho}). Whereas for the concentrations $0.27\leq x\leq0.58$ and $x=1$, the AFM ordering results in a drop of $\rho(\rm{T})$, for $0.18\leq x\leq0.215$ and $x=0.68$ we find an increase of $\rho(\rm{T})$ just below $T_{\rm{N}}$. This upturn has already been observed for $x=0.07$ (Ref.~\cite{Friedemann:2009}) and 0.12 (Ref.~\cite{Friedemann_resistivity_09}) and probably emerges from a gap opening at the Fermi level below $T_{\rm{N}}$ in the new AFM Brillouin zone. The second anomaly $T_{\rm{L}}$ for concentrations $0.18\leq x\leq0.215$ is again marked by a decreasing $\rho(T)$. While this feature disappears with increasing Co concentration, it is recovered for $x\geq 0.68$.

Analyzing the resistivity data in the vicinity of the magnetic transitions, another important fact can be seen in the inset of Fig.~\ref{FigRho}. Comparing the low-temperature behavior of $\rho(\rm{T})$ for $x=0.215$ and 0.27, a pronounced change from a small increase to a distinct drop in $\rho(T)$ at $T_{\rm{N}}$ can be noticed. This different temperature dependence is most likely due to a change of the magnetic structure between these two concentrations. Finally, we have to point out that for $0.18\leq x\leq0.215$, the resistivity is quasi-linear in $T$ above the ordering temperature, as already shown for YbRh$_2$Si$_2$,\cite{trovarelli_ybrh2si2_2000} whereas for $x=0.27$ and $x=0.38$ it is of a sublinear type (see inset of Fig.~\ref{FigRho}). A detailed analysis of the resistivity down to 20\,mK will be published elsewhere. \cite{Lausberg_private}

\subsection{\label{SpecificHeat}Specific heat}
\begin{figure}%
\includegraphics[width=\columnwidth]{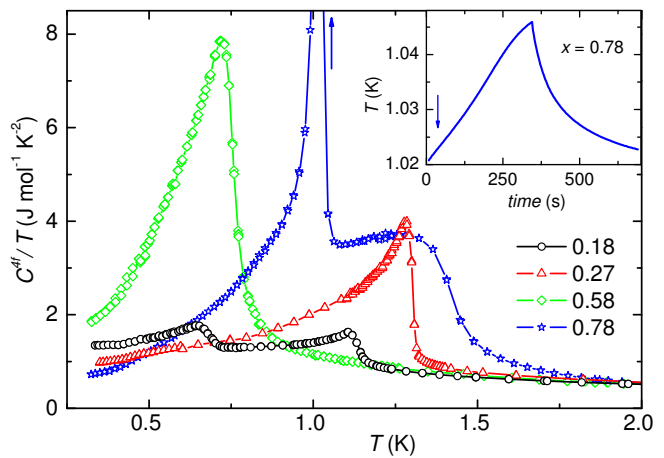}%
\caption{\label{FigC}(Color online) $4f$ contribution to the specific heat for $x=0.18$, $0.27$, $0.58$, and $0.78$ evidencing different behaviors at the magnetic phase transitions in Yb(Rh$_{1-x}$Co$_x$)$_2$Si$_2$. For $x=0.78$, the corresponding relaxation curve at the first-order phase transition temperature, $T_{\rm{L}}=1.01$\,K is plotted in the inset, with the arrow marking the onset of the latent-heat contribution.}%
\end{figure}

Now, we present the $4f$ contribution to the specific heat, $C^{4f}$, for the single crystals with $0.18\leq x\leq 1$ at temperatures $0.35\leq T\leq5$\,K measured by the relaxation method in the PPMS in zero magnetic field. The lattice contribution has been subtracted from the total specific heat using the data for the non-magnetic reference compound LuRh$_2$Si$_2$.\cite{ferstl_new_2007} Since the lattice contribution at 1\,K amounts to less than 1\% of the total measured specific heat, the synthesis and measurement of the appropriate reference systems Lu(Rh$_{1-x}$Co$_x$)$_2$Si$_2$ was not necessary. 

In Fig.~\ref{FigC}, $C^{4f}/T$ vs. $T$ for the concentrations $x=0.18$, $0.27$, $0.58$, and $0.78$ is depicted. For $x=0.18$, we detect two mean-field-type anomalies at $T_{\rm{N}}=1.10\pm0.03$\,K and $T_{\rm{L}}=0.65\pm0.03$\,K. Below $T_{\rm{L}}$, the Sommerfeld coefficient $\gamma_0=1.3\,\rm{J}/\rm{mol K}^2$ is still very high. Raising the cobalt content to $x=0.27$, the electronic specific heat changes significantly. For this amount of doping, $C^{4f}/T$ shows only one $\lambda$-type phase transition at $T_{\rm{N}}=1.29\pm0.03$\,K. The peak is much more pronounced and sharp, compared to the broad mean-field-type anomalies for $x=0.18$, which indicates, that the substitution of cobalt does not lead to significant disorder effects which would further smear the transition anomalies. The analysis of the critical magnetic fluctuations around $T_N$  reveals a conventional critical exponent for $x=0.27$ and 0.38; \cite{Krellner:2010} whereas, for $x\leq 0.215$ such an analysis could not be carried out due to the broader mean-field shape of the phase transitions. 

With increasing Co content the behavior of the specific heat changes again. For $x=0.58$, the AFM ordering temperature $T_{\rm{N}}=0.73\pm0.03$\,K assumes a lower value, compared to the one for $x=0.38$ ($T_{\rm{N}}=1.22$\,K), breaking the monotonic increase of $T_{\rm{N}}$. This decrease of $T_{\rm{N}}$ between $x=0.38$ and $0.58$ will be discussed below. Further on, the absolute value of $C^{4f}/T$ at $T_{\rm{N}}$ is enhanced in comparison to the absolute values of the other concentrations and also the value at the lowest accessible temperature of $T=0.35$\,K is considerably enlarged, which might be caused by another successive magnetic phase transition at lower temperatures. 

Two magnetic phase transitions are again visible for high concentrations of cobalt, e.g. $x=0.78$, with $T_{\rm{N}}=1.32\pm0.03$\,K and $T_{\rm{L}}=1.01\pm0.03$\,K, as shown in Fig.~\ref{FigC}. The difference to the low-doped samples, e.g. $x=0.18$, is that $T_{\rm{L}}$ shows clear signatures of a first-order phase transition. The values of $C(T)$ across the sharp peak at $T_{L}$ where taken continuously by calculating the time derivative of the relaxation function plotted in the inset of Fig.~\ref{FigC}.\cite{lashley_critical_2003} The non-exponential temperature increase in time in the heating part is a direct evidence for a first-order phase transition. For $x=0.78$, the upper transition at $T_{\rm{N}}=1.32$\,K is a second-order mean-field-type AFM phase transition. It is, however, less sharp, probably owing to increasing disorder effects for large amounts of cobalt substitution. Furthermore, the Sommerfeld coefficient appears to be substantially smaller in comparison to the lower doped samples. 

\begin{figure}%
\includegraphics[width=\columnwidth]{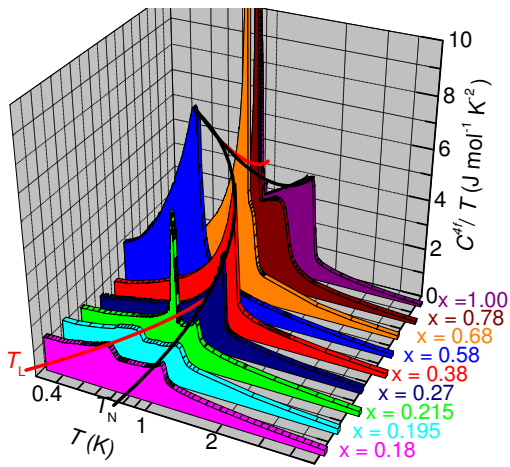}%
\caption{\label{FigCall}(Color online) $4f$ contribution to the specific heat for all concentrations $x\geq0.18$. The black and red lines mark the magnetic phase transitions $T_{\rm{N}}$ and $T_{\rm{L}}$, respectively.}%
\label{HeatCapacity2}%
\end{figure}

The heat-capacity measurements of all doped samples are depicted in a three-dimensional representation in  Fig.~\ref{FigCall}, which displays the complexity of the magnetic behavior in this series. All extracted transition temperatures are listed in Tab.~\ref{TabCW}. Although several details regarding the magnetic properties of Yb(Rh$_{1-x}$Co$_x$)$_2$Si$_2$ are not completely understood, a global picture may be drawn. The obtained results can be divided into three different regions depending on the amount of cobalt substitution. (1) For $0.18\leq x\leq 0.215$ two phase transition temperatures are clearly visible. With increasing Co concentrations, $T_{\rm{N}}$ is almost constant whereas $T_{\rm{L}}$ steadily increases towards higher values. In addition, the phase transition at  $T_{\rm{L}}$ is changing its nature between $x=0.195$ and $x=0.215$ from second to first order. (2) Above $x=0.215$, $T_{\rm{L}}$ merges the $T_{\rm{N}}$-line, since for $0.27\leq x\leq0.58$ only one transition temperature is observed. This phase transition is of second order, with $T_{\rm{N}}$ for $x=0.58$ being reduced compared to the value at lower Co concentrations. (3) Above $x=0.58$, a $T_{\rm{L}}$-line again emerges from $T_{\rm{N}}$. In contrast to those samples with lower Co content the transition at $T_{\rm{L}}$ is now always of first order and shifts towards lower temperatures with increasing Co doping. On the other hand, a second-order phase transition takes place at $T_{\rm{N}}$ which increases upon increasing substitution reaching a value of $1.61\pm0.03$\,K for $x=1$. Preliminary specific-heat measurements with applied magnetic field for selected concentrations show a shift of $T_N$ and $T_L$ to lower temperatures with increasing magnetic field confirming the antiferromagnetic nature of the phase transitions.\cite{Klingner:2010}

\subsection{\label{PES}Photoemission spectroscopy}
The results of earlier measurements on YbCo$_2$Si$_2$ performed by means of M\"ossbauer\cite{hodges_magnetic_1987} and magnetic susceptibility\cite{kolenda_valence_1989} techniques suggest that Yb reveals a purely trivalent state. On the other hand, for YbRh$_2$Si$_2$ a slight deviation from trivalent Yb was found with a mean valence of $2.9$.\cite{PRB_Danzenbaecher2007} Hence, it is assumed that a gradual valence change can be followed when going from $x=0$ to 1. To this end, we applied X-ray photoemission spectroscopy (PES) to a series of single crystals with $x=0$, $0.07$, $0.12$, $0.38$, $0.68$, and $1$. PES addresses the electronic states under consideration directly and is thus an excellent probe to study the mean valence of a compound.\cite{PRB_Danzenbaecher2007, Moreschini_PRB2007} The experiments were performed at the SPring-8 facility using synchrotron radiation delivered by beamline BL25SU on single crystals cleaved immediately before the measurements, at $23$\,K and a pressure around  $2 \cdot 10^{-10}$\,mbar.

\begin{figure}%
\includegraphics[width=\columnwidth]{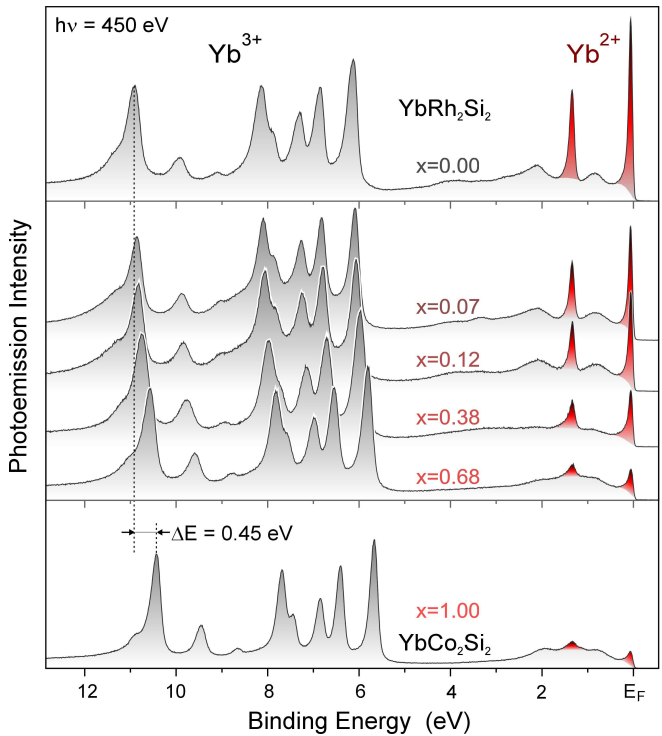}%
\caption{\label{FigPES}(Color online) Photoemission spectroscopy on cleaved single crystals for different $x$ showing the gradual disappearance of the intensity of the bulk divalent components close to $E_F$ with increasing doping.}%
\end{figure}

Fig.~\ref{FigPES} presents the valence-band angle-integrated PES spectra taken at $450$\,eV photon energy for a series of Yb(Rh$_{1-x}$Co$_x$)$_2$Si$_2$ single crystals. At this photon energy PES can already probe deeper than the first few surface layers down to $\sim 10$\,\AA,\cite{Seah:1979, Suga:2005, Shigemoto:2007} while the total energy resolution of the instrument still remains reasonably high. The spectra exhibit two prominent regions, one between $\sim1.3$\,eV binding energy (BE) and the Fermi level ($E_F$) emphasized by red, and a second one at $5-12$\,eV BE shaded light-gray. The spin-orbit split sharp doublet near the Fermi level represents the $4f$ electron emission from divalent Yb atoms, while the more complex spectral structure at higher binding energies reflects the trivalent Yb state.\cite{Gerken_1983} The observed large energy shift between the divalent and trivalent Yb $4f$ components is explained by the large Coulomb-correlation energy in the final state of photoemission. All spectra were taken at Si-terminated surface regions of the \textit{in situ} cleaved samples. Therefore, contributions of purely divalent surface Yb atoms (broad doublet with components at $0.8$ and $2.0$\,eV BE) are largely suppressed.

The PES spectrum taken from pure YbRh$_2$Si$_2$ exhibits both divalent and trivalent $4f$ components, typical for an intermediate-valent ground state of this material, with the possibility of strong hybridization between $4f$ and valence states, which was indeed observed in our previous experiments.\cite{PRB_Danzenbaecher2007,vyalikh_photoemission_2008} Upon gradual substitution of Rh atoms by Co, the intensity of the bulk divalent components close to $E_F$ decreases considerably, in comparison to the trivalent features and almost disappears for the pure YbCo$_2$Si$_2$ compound. Moreover, the trivalent multiplet experiences a rigid shift towards the Fermi level by $0.45$\,eV. Reason for that might be a different chemical shift of the trivalent Yb component for YbRh$_2$Si$_2$ and YbCo$_2$Si$_2$. In fact, thermo-chemical calculations of the Yb$^{3+}$ energy shift for both compounds\cite{Laubschat:1986} propose a shift to lower BE when going from the Rh to the Co compound. Moreover, the decrease of the divalent Yb component when going from $x=0$ to 1 shows that the respective state in the ground state shifts to higher energies above $E_F$. Assuming that the Coulomb-correlation energy in the photoemission final state is comparable for both compounds the trivalent Yb component should shift towards $E_F$ accordingly. Which of both effects is dominating here remains subject of further investigations. In any case, the experimental observations clearly reveal the stabilization of the trivalent state of Yb ions in pure YbCo$_2$Si$_2$ which would characterize this material as showing a rather weak interaction between the $4f$ and the valence states in comparison to YbRh$_2$Si$_2$. A thorough analysis of the mean Yb valency for the different concentrations based upon results of high-resolution resonant inelastic X-ray scattering experiments, is presently underway.\cite{Kummer:2010}

\section{\label{Discussion}Discussion} 

\subsection{\label{MagneticPhaseDiagram}Magnetic phase diagram Yb(Rh$_{1-x}$Co$_x$)$_2$Si$_2$}

\begin{figure}%
\includegraphics[width=\columnwidth]{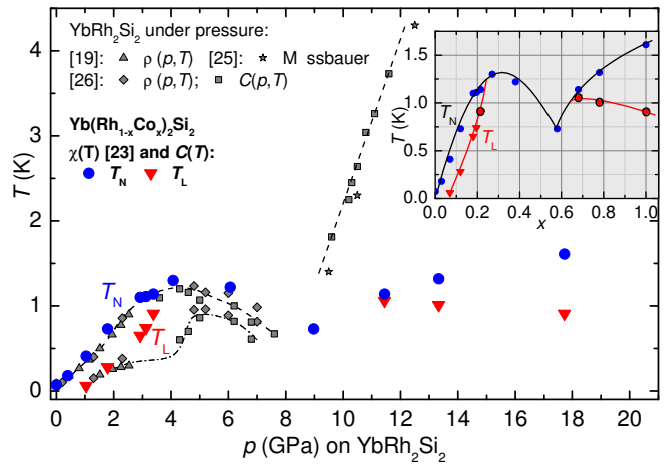}%
\caption{\label{FigPD}(Color online) Magnetic phase diagram of the series Yb(Rh$_{1-x}$Co$_x$)$_2$Si$_2$ in comparison to the  results from pressure experiments reported in the literature. \cite{Mederle:2002,Plessel:2003,Knebel:2006} Clearly visible is the good agreement of the data for the low doped region. For high amounts of cobalt ($x\geq 0.68$) $T_N$ increases much faster under pressure than for Co substitution.}%
\end{figure}

The results obtained by magnetic, transport, thermodynamic, and spectroscopic measurements lead to a consistent picture of the magnetic behavior of the Yb(Rh$_{1-x}$Co$_x$)$_2$Si$_2$ single crystals. The $x-T$ phase diagram is presented in the inset of Fig.~\ref{FigPD}. For the sake of simplicity, the transition temperatures $T_{\rm{N}}$ and $T_{\rm{L}}$ are presented for the heat-capacity measurements only. For $x\leq 0.12$, the results from low-temperature susceptibility data are taken from the literature.\cite{Westerkamp:2008}   First-order phase transitions at $T_{\rm{L}}$ for $x=0.215$ as well as for $0.68\leq x\leq 1$ are marked by black circles. The transition temperatures for $x\geq  0.18$ obtained from resistivity and susceptibility measurements are in very good agreement with these specific-heat results. Since the analogy between hydrostatic and chemical pressure effects is one of the interesting aspects of this work, we compare in the main part of Fig.~\ref{FigPD} our results for the alloy with the reported results for pure YbRh$_2$Si$_2$ under pressure. By using the measured lattice parameters (see Tab.~\ref{TabXRD}) and the bulk modulus of YbRh$_2$Si$_2$ ($189$\,GPa),\cite{Plessel:2003} we can convert the Co concentration into chemical pressure.

We shall first address the phase diagram of the alloy series on its own, and then discuss the comparison with hydrostatic pressure results. The main features in the $x-T$ magnetic phase diagram of the alloy are a steep increase of $T_N(x)$ with $x$ at low Co contents, a broad maximum near $x=0.3$ with $T_{\rm{N,max}} = 1.30$\,K, a decrease towards a minimum near $x=0.58$ with $T_{\rm{N,min}}=0.73$\,K, followed by a further increase towards $T_N = 1.61$\,K in pure YbCo$_2$Si$_2$. The transition at $T_N$ seems to be second order in the whole concentration range. Further on, a second transition at $T_L < T_N$ appears near $x = 0.07$, shifts continuously to higher temperatures with increasing $x$, changes from second-order to first-order type around $x = 0.21$, and finally merges with $T_N(x)$ near $x = 0.27$. A second transition at $T_L < T_N$, of first order type, is also observed at higher Co contents $x > 0.58$, beyond the minimum in $T_N(x)$. Its critical temperature $T_L$ is almost insensitive to the Co content. Our results indicate that in the whole concentration range the anomaly at $T_N$ corresponds to a transition towards an AFM state, despite the Weiss temperature becoming FM in the region $0.2 < x < 0.5$. We have no evidence for a change towards FM ordering. On the other hand, the anomalies at $T_L$ for low or high Co contents are obviously connected with a change in the AFM structure, i.e., in the propagation vector and/or in the orientation of the Yb moments. This phase diagram is rather complex, with a pronounced non-monotonic behavior, indicating a competition between different interactions: e.g., Kondo effect, different exchange paths, and CEF effects.

It is therefore quite amazing that the main features of this complex phase diagram are nicely reproduced in the pressure experiments. The non-monotonous volume dependence of $T_N$ is almost identical for Co substitution and hydrostatic pressure up to $x=0.58$ or $p=8$\,GPa. The appearance and initial increase of $T_L$ is also identical in pressure and Co substitution experiments, but the further evolution is different. In the alloy, $T_L$ merges with $T_N$ near $x = 0.27$, and we did not found any evidence for a second transition below $T_N$ in the range $0.27 \leq  x \leq  0.58$. In contrast, in the pressure experiments of Knebel \textit{et al.}\cite{Knebel:2006}, the second transition could not be resolved in the pressure range 2.3 $< p < 4.6$\,GPa, but was again clearly observed in specific heat data at higher pressures, 4.6$\leq  p \leq  6.8$\,GPa. There, the transition temperature $T_L(p)$ first increases sharply and then stays slightly below $T_N(p)$.

The most pronounced differences are however observed beyond $x = 0.58$ or $p = 8$\,GPa. In diamond anvil cell experiments a sharp increase in $T_N$ with $p$, up to $T_N \approx 7.5$\,K is observed.\cite{Plessel:2003} In contrast, in the alloy series we found a much weaker increase of $T_N$, only up to 1.6\,K in pure YbCo$_2$Si$_2$, and a further transition at $T_L < T_N$ for $x \geq  0.68$. At least two effects can be proposed as origin of this clear difference: (1) The increase of $c/a$ upon Co doping might result in a significant difference between the CEF under hydrostatic and chemical pressure at higher $p$. In this context one should note that the M\"ossbauer experiments in the diamond anvil cell indicate a switch of the direction of the ordered moment from the suspected in-plane direction at low $p$ to along $c$ for $p > 9$\,GPa, while M\"ossbauer results on pure YbCo$_2$Si$_2$ indicate the ordered Yb moment to be within the basal plane.\cite{hodges_magnetic_1987} (2) For hydrostatic pressure only a small part of the volume reduction at $p = 10$\,GPa can be attributed to the shrinking of the Yb atomic volume connected with the slight reduction of the Yb valence from $\approx 2.9$ at $p = 0$ to pure trivalent at high pressures. The largest part of the volume reduction at higher pressures corresponds to a real compression of Rh, Si, and Yb atoms which should result in an increase of hybridization and thus an increase of the RKKY interactions.

\subsection{\label{KondoRKKY} Kondo versus RKKY exchange interaction}
In the previous paragraph, the evolution of the magnetic behavior of Yb(Rh$_{1-x}$Co$_x$)$_2$Si$_2$ has been related to the change of the average unit-cell volume, $V$. One of the important mechanism for the understanding of the phase diagram is the competition between the local Kondo and the inter-site RKKY exchange interactions. This competition is directly connected to the strength of the hybridization between the localized $4f$ states and the delocalized valence-band states, expressed by the exchange constant $J$ which is a function of $V$.

For the heavy-fermion metal YbRh$_2$Si$_2$, with a large Sommerfeld coefficient ($\gamma_0=1.7$\,J/molK$^2$) this hybridization is strong and the Kondo effect dominates over the RKKY interaction. A Kondo temperature for the lowest-lying CEF Kramers doublet, $T_{\rm{K}}\approx25$\,K, has been deduced from the magnetic entropy.\cite{trovarelli_ybrh2si2_2000, Gegenwart:2006} For Yb compounds, the application of pressure leads to a stabilization of the magnetic, trivalent Yb state, which in a simple Kondo-lattice model corresponds to a decrease of $J$. Due to the different analytical dependencies of the energy scales $T_{\rm{K}}$ and $T_{\rm{RKKY}}$ on the hybridization strength $J$, this decreasing $J$ results in a much faster exponential decrease of $T_{\rm{K}}$ compared with the quadratic decrease of the RKKY energy scale.\cite{doniach_kondo_1977}  Hence, at some point the RKKY interaction becomes dominant, and the AFM ordering is stabilized. 

\begin{figure}%
\includegraphics[width=\columnwidth]{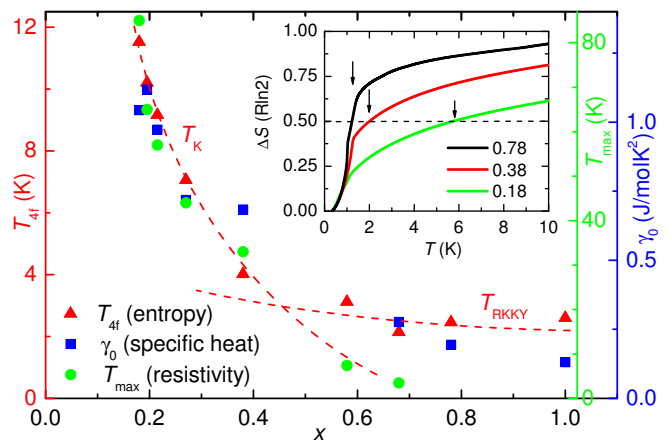}%
\caption{\label{FigTK}(Color online) Evolution of the characteristic energy scales of Yb(Rh$_{1-x}$Co$_x$)$_2$Si$_2$ illustrated by the parameters $T_{4\rm{f}}$, $T_{\rm{max}}$, and $\gamma_0$ evidencing the change of the dominant energy scale from Kondo to RKKY (dashed lines are guide to the eyes). The inset shows the increment of the entropy, $\Delta S$, used to determine $T_{4\rm{f}}$.}%
\end{figure}

The decrease of $T_K$ and the change from a Kondo- to a RKKY-dominated behavior in Yb(Rh$_{1-x}$Co$_x$)$_2$Si$_2$ is illustrated in Fig.~\ref{FigTK} by plotting several parameters such as $T_{4\rm{f}}$, the Sommerfeld coefficient $\gamma_0$, and the position of the high-temperature maximum in $\rho(T)$, $T_{\rm{max}}$, as function of $x$. The parameter $T_{4\rm{f}}$ (triangles in Fig.~\ref{FigTK}) reflects the energy scale of all exchange interactions acting on the lowest CEF doublet and has been deduced from the magnetic entropy, $S$, as $T_{4\rm{f}}=2\cdot T (S=0.5R\rm{ln}\,2)$ (see arrows in the inset of Fig.~\ref{FigTK} at $S=0.5R\rm{ln}\,2$). For $x\leq 0.38$, $T_{4\rm{f}}$ is far above $T_{\rm{N}}$, highlighting a dominating Kondo energy scale. With increasing amount of cobalt $T_{4\rm{f}}$ strongly decreases reflecting the decrease of $T_K$. However, the strong decrease of $T_{4\rm{f}}$ stops for $x>0.4$ and is replaced by an almost $x$-independent behavior. Since this change occurs where $T_{4\rm{f}}$ is approaching $T_N$, it obviously reflects the crossover of the two energy scales $T_K$ and $T_{\rm{RKKY}}$, the later one becoming the dominant one for $x>0.6$. This overall behavior is confirmed by the evolution of the coherence temperature $T_{\rm{max}}$ (circles in Fig.~\ref{FigTK}), deduced from the maximum in $\rho(T)$ (cf. Sec.~\ref{SecResistivity}). At low Co concentrations, $T_{\rm{max}}$ decreases strongly with increasing $x$, showing an almost perfect scaling with $T_{4\rm{f}}$. For $x>0.4$, $T_{\rm{max}}$ proceeds decreasing, in contrast to $T_{4\rm{f}}$, directly evidencing a further reduction of $T_K$. For $x=1$, $T_{\rm{max}}$ shifts below $T_{\rm{N}}$, which implies $T_K$ well below $T_N$. The decrease of $T_{\rm{K}}$ is nicely confirmed by the increasing effective moments at low temperatures, $\mu_{\rm{eff}}^{\rm{ab};2-4\rm{K}}$ (see Tab.~\ref{TabCW} and Fig.~\ref{FigCW}), which reflects the diminished screening of the local $4f$ moments upon increasing Co content. The Sommerfeld coefficient $\gamma_0$ (squares in Fig.~\ref{FigTK}), which was deduced from a linear fit of  $C^{4f}/T$ versus $T^2$ below the phase transition temperatures, assuming a dominant contribution of AFM magnons to the specific heat, also decreases continuously with increasing Co content, from the huge value $\gamma_0=1.7\,\rm{J/molK}^2$ in pure YbRh$_2$Si$_2$ to merely $\gamma_0=0.13\pm0.05\,\rm{J/molK}^2$ in pure YbCo$_2$Si$_2$. The decrease of $\gamma_0$ is smoother than the decrease of $T_K$. From a very simple phenomenological approach one expects a decrease of $\gamma_0$ with $T_K$ in the magnetic ordered region of the phase diagram of a Kondo lattice, but there is no strict theoretical prediction, and the relation is not universal, since $\gamma_0$ depends not only on $T_K$, but also on the molecular field.\cite{Bredl:1978} Further support for the decrease of the quasiparticle mass comes from the PES results for different $x$ (see Fig.~\ref{FigPES}). The strong reduction of the density of states at the Fermi level due to a drop of the Yb$^{2+}$ peak at $E_F$, is directly visible.

Finally, we discuss the $4f$ contribution to the entropy which was obtained by integrating $C^{4f}/T$ over $T$ from the lowest measured temperature $T=0.35$\,K (cf. inset of Fig.~\ref{FigTK}). Two important features can be deduced from the behavior of $S(T)$ under increasing chemical pressure. First, since the entropy does not exceed $R\rm{ln}\,2$ at $T=10$\,K throughout the whole series, only the lowest CEF doublet is involved in the formation of the ground state, the first excited CEF doublet remains energetically well separated. Second, the smooth and continuous  increase of $S$ with increasing amount of cobalt rules out an abrupt change from a low moment to a high moment state, but rather demonstrates a smooth stabilization of the AFM order. This behavior is evident from $S(T)$ at $T_{\rm{N}}$. For low $x$, $S(T_{\rm{N}})$ is only a small fraction of $R\,\rm{ln}2$, e.g., $S(x=0.18, T=T_N)\sim0.3R\,\rm{ln}2$, but with increasing Co concentrations, $S(T_N)$ rises  continuously  towards higher values, e.g., $S(x=0.78, T=T_N)\sim0.7R\,\rm{ln}2$.

\section{Conclusion}
We have reported a detailed investigation of the evolution of magnetism in the alloy Yb(Rh$_{1-x}$Co$_x$)$_2$Si$_2$. We first described the crystal growth and the structural and chemical characterization, and then presented measurements of the magnetic susceptibility, electrical resistivity, specific heat, and photoemission for different concentrations covering the whole concentration range. The results revealed a complex magnetic phase diagram, with an antiferromagnetic ground state at all compositions, but with a non-monotonic evolution of the N\'eel temperatures $T_N$ and a second antiferromagnetic transition at $T_L<T_N$ for some concentrations. $T_N$ first increases with increasing $x$, from 72 mK in pure YbRh$_2$Si$_2$ to a first maximum with $T_N =1.3$\,K near $x=0.3$, as expected for an Yb system under chemical pressure. However, for $x > 0.3$, instead of increasing further, $T_N$ decreases smoothly to $T_N = 0.73$\,K near $x = 0.58$, and rises again smoothly to $T_N = 1.6$\,K in pure YbCo$_2$Si$_2$. Some significant changes in the magnetic state seem to occur in the regions $x \approx 0.3$ and $x \approx 0.6$, since they not only correspond to extrema in $T_N(x)$, but also to tricritical points where a second transition line $T_L(x)$ merges with $T_N(x)$. The $T_L$ lines extend from the two tricritical points at $x \approx 0.27$ and $x \approx 0.58$ to almost pure YbRh$_2$Si$_2$ and pure YbCo$_2$Si$_2$, respectively. The anomalies at $T_L$ at low or high Co contents are obviously connected with a change in the  antiferromagnetic structure. While the transition at $T_N$ is always second-order type, the transition at $T_L$ is of first-order type in the whole  region at high Co contents but switches from first-order to second-order type when moving away from the tricritical point at low Co contents. Despite its complexity, this phase diagram is very similar to the phase diagram under hydrostatic pressure when both are plotted as a function of the unit cell volume. For small reduction of the volume they are identical, while some differences appear at larger volume reduction, the most prominent being a strong increase of $T_N$ for $p > 8$\,GPa contrasting a much smaller increase for large Co substitution. The strong similarity indicates that Co substitution basically corresponds to chemical pressure with no evidence for Co magnetism. Furthermore, analysis of susceptibility, resistivity and specific heat data indicates a pronounced decrease of the Kondo temperature with increasing Co content. For $x > 0.38$, the Kondo temperature becomes comparable to $T_{\rm{N}}$, resulting in a crossover from itinerant to localized Yb-$4f$ states as well as a change of the dominant mechanism from the local Kondo to the inter-site RKKY interaction. This localization of the $4f$ electrons is accompanied by drastic changes in the low-temperature properties studied by magnetization, specific heat, resistivity, and photoemission spectroscopy. In addition, the evolution of the susceptibility data in the series confirms a competition between ferromagnetic and antiferromagnetic exchange as first suggested from NMR data in pure YbRh$_2$Si$_2$.\cite{Ishida:2003} For $0.2 < x < 0.5$ the Weiss temperature determined from Curie-Weiss plots at low $T$ becomes ferromagnetic, although ordering stays antiferromagnetic. Therefore, the series Yb(Rh$_{1-x}$Co$_x$)$_2$Si$_2$ allows direct experimental access to the two competing magnetic correlations in YbRh$_2$Si$_2$ at finite and zero wave vector and offers a unique opportunity to get a deeper understanding of the nearby unconventional quantum critical point. For example, tracing the magnetic propagation vector from YbCo$_2$Si$_2$ to YbRh$_2$Si$_2$ using neutron diffractometry may be one key experiment to clarify the magnetic structure of YbRh$_2$Si$_2$.

\section*{Acknowledgements}
The authors thank U. Burkhardt and P. Scheppan for energy dispersive X-ray analysis of the samples as well as C. Klausnitzer and R. Weise for technical assistance. 
The photoemission experiment at SPring-8 was performed with the approval of Japan Synchrotron Radiation Research Institute (Proposal No. 2009A1029).
We acknowledge valuable discussions with
M. Baenitz,
S. Friedemann, 
P. Gegenwart,
T. Gruner,
A. Haase, 
A. Jesche,
K. Kaneko,
S. Kirchner, 
S. Lausberg 
N. Mufti,
L. Pedrero, 
Q. Si,
J. Sichelschmidt, 
A. Steppke, 
O. Stockert,
T. Westerkamp,
S. Wirth,
and 
J. Wykhoff 
.
The DFG (Research Unit 960, "Quantum phase transitions") is acknowledged for financial support.

\end{document}